\documentclass[fleqn,twoside]{article}
\usepackage{espcrc2}

\usepackage{graphicx}
\usepackage[figuresright]{rotating}

\newcommand{\AmS}{{\protect\the\textfont2
  A\kern-.1667em\lower.5ex\hbox{M}\kern-.125emS}}
\newcommand{\mpi}{m_{\pi}}
\newcommand{\non}{\nonumber}
\newcommand{\bes}{\begin{displaymath}}
\newcommand{\ees}{\end{displaymath}}
\newcommand{\be}{\begin{equation}}
\newcommand{\ee}{\end{equation}}
\newcommand{\beas}{\begin{eqnarray*}}
\newcommand{\eeas}{\end{eqnarray*}}
\newcommand{\bea}{\begin{eqnarray}}
\newcommand{\eea}{\end{eqnarray}}

\title{Hadron Mass Extraction from Lattice QCD}

\author{S. V. Wright\address[Adl]{Department of Physics and
        Mathematical Physics\\
	and Special Research Centre for the Subatomic Structure of
        Matter,\\ 
	University of Adelaide, Adelaide 5005, Australia}%
	\thanks{Present address: Division of Theoretical Physics,
	Department of Mathematical Sciences, University of Liverpool,
	Liverpool L69 3BX, UK},
	D. B. Leinweber\addressmark,
	A. W. Thomas\addressmark 
	{ }and{ }  
	K. Tsushima\addressmark%
	\thanks{Present address: Department of Physics and Astronomy,
	University of Georgia, Athens, GA 30602, USA}}

\begin{document}

\begin{abstract}

\vspace{-4.cm}%
\hfill $\begin{array}{r@{\mathrm{-}}l}
\mathrm{ADP} & \mathrm{01-48/T480}\\
\mathrm{LTH} & \mathrm{526}
\end{array}$
\vspace{3.5cm}%

The extraction of quantities from lattice QCD calculations at
realistic quark masses is of considerable importance.  Whilst physical
quark masses are some way off, the recent advances in the calculation
of hadron masses within full QCD now invite improved extrapolation
methods.  We show that, provided the correct chiral behaviour of QCD
is respected in the extrapolation to realistic quark masses, one can
indeed obtain a fairly reliable determination of masses, the sigma
commutator and the $J$ parameter.  We summarise these findings by
presenting the nonanalytic behaviour of nucleon and rho masses in the
standard Edinburgh plot.

\vspace{1pc}
\end{abstract}

\maketitle

\section{INTRODUCTION}

There are well-known difficulties associated with making dynamical
fermion lattice QCD calculations at light quark masses.  There is the
need however, to relate quantities calculated on the lattice with
physical observables, hence results are required at physical quark
masses.  These two mutually exclusive restrictions on
the field have motivated the necessity for extrapolation from the
region in which calculations are able to be performed --- that is, the
region of unphysically heavy quarks --- to lighter masses, including the
physical quark masses.  In this paper we discuss the construction and
application of an extrapolation method for masses
\cite{Leinweber:2000ig,Leinweber:2001ac} that respects the
correct chiral behaviour of QCD and also allows the extraction of
other quantities \cite{Leinweber:2001ac,Leinweber:2000sa}.  This
approach is not limited to the case of masses
in dynamical fermion lattice QCD calculations.  Other successes of
this approach may be found, for example, in the extrapolation of 
baryon charge radii \cite{Hackett-Jones:2000js}, magnetic moments
\cite{Hackett-Jones:2000qk}, structure functions 
\cite{Detmold:2001jb,Detmold:2001dv} and quenched QCD data \cite{Ross}. 

\subsection{Goldstone Boson Loops}

It is accepted that Goldstone Boson loops play an important role in
all hadronic properties --- their role is in one sense the basis
of Chiral Perturbation Theory ($\chi$PT).  Lattice QCD calculations,
as an ab initio approach to calculating quantities in QCD, implicitly
includes these loop contributions.  It has become clear
recently, with calculations appearing at lighter quark masses
\cite{Allton:1999gi,Aoki:1999ff}, that the na\"{\i}ve linear
extrapolation methods are not reproducing the data.  In particular in
\cite{Aoki:1999ff} it was stated
\begin{quotation}
``The existence of curvature [at small quark masses] is observed,
necessitating a cubic Ansatz for extrapolation to the chiral limit.''
\end{quotation}

The following section reviews how the inclusion of chiral physics
allows reliable extrapolations of lattice QCD calculations
\cite{Leinweber:2000ig,Leinweber:2001ac}.  Section 3 reports new
results for the Edinburgh plot.

\section{EXTRAPOLATION METHODS}

In QCD chiral symmetry is dynamically broken, and the pion is almost a
Goldstone boson. It plays a significant role in the self-energy
contributions to the $N$ and $\Delta$, because of the strong
coupling to the baryons.  Chiral symmetry requires that, in the region
where perturbations around light quarks makes sense, the mass of the
nucleon has the form
\bea
m_{N}(\mpi) & = & m_{N}^{(0)} + \alpha \mpi^{2} + \beta \mpi^{3} + \non\\
            &   & \gamma \mpi^{4} \ln \mpi + \ldots \, ,
\eea
where $m_{N}^{(0)}$, $\alpha$, $\beta$ and $\gamma$ are functions of
the strong coupling constant.  In particular the values of the
coefficients of the non-analytic (in quark mass) terms --- recall
that $\mpi^{2} \propto m_{q}$ --- are known exactly from
$\chi$PT\@. However it is only recent results from the lattice that have
indicated any need of higher order terms beyond that of a linear 
extrapolation in quark mass (or $\mpi^{2}$).

\subsection{Chirally Motivated Form}

An attempt at having a chirally motivated form for extrapolating
masses has been \be m_{N}(\mpi) = m_{0} + \tilde{\alpha} \mpi^{2} +
\tilde{\beta} \mpi^{3} \, , \label{eqn:cubic} \ee where $m_{0}$,
$\tilde{\alpha}$ and $\tilde{\beta}$ are fit parameters.  Na\"{\i}vely
this is a good choice.  It reflects the known non-analyticity from
$\chi$PT and still reproduces the lattice results.  The problem with
this method is associated with the choice of $\tilde{\beta}$.  The
value of the coefficient of the cubic term is known explicitly in
$\chi$PT\@.  So a functional form, motivated by chiral symmetry,
should preserve the known value of $\tilde{\beta}$.  Optimising
$\tilde{\beta}$ via a best fit to existing lattice data provides
$-0.55$ GeV$^{-2}$.  However, the result from $\chi$PT is $-5.6$
GeV$^{-2}$.  That the coefficient is so small is not surprising.  The
functional form attempts to reproduce the lattice data over a large
range of $\mpi^{2}$, where the data is predominantly linear --- as can
be seen in Fig.~\ref{fig:Masses}.  However $\chi$PT is an expansion
about the massless quark limit and would not be expected to be
applicable (or even convergent) at such large quark masses.

\subsection{Current Calculation}

It has been found \cite{Leinweber:2000ig,Leinweber:2001ac} that by
retaining the contributions to the self-energy of the hadron mass that
vary the most rapidly with $\mpi$ near the chiral limit, a successful
extrapolation method may be formed.  This methodology includes the
most important non-analytic structure in the hadron mass near the
chiral limit with {\em exactly\/} the correct coefficients.  The pion
mass dependence of the masses of the $N$, $\Delta$ and $\rho$ are:
\bea
m_{N} & = & a_{0} + a_{2} \mpi^{2}
      + \sigma_{N N\pi}(\Lambda_{N},\mpi) \non \\
& & + \sigma_{N \Delta\pi}(\Lambda_{N},\mpi) \label{eqn:N}\\
m_{\Delta} & = & b_{0} + b_{2} \mpi^{2}
      + \sigma_{\Delta\Delta\pi}(\Lambda_{\Delta},\mpi) \non \\
& & + \sigma_{\Delta N\pi}(\Lambda_{\Delta},\mpi) \label{eqn:Delta}\\
m_{\rho} & = & c_{0} + c_{2} \mpi^{2}
      + \sigma_{\rho\omega\pi}(\Lambda_{\rho},\mpi) \non \\
& & + \sigma_{\rho\pi\pi}(\Lambda_{\rho},\mpi) \label{eqn:rho}
\eea
where $\sigma_{ABC}$ indicates the contribution from the
$A \to BC \to A$ self-energy process.  The expressions
for these self-energy contributions for the $N$ and $\Delta$ may be
found in \cite{Leinweber:2000ig}.  The two significant processes for
the $\rho$ are the $\rho\to\omega\pi$ and $\rho\to\pi\pi$
self-energies and they are presented in \cite{Leinweber:2001ac}.

An additional level of detail explicitly included in these
extrapolation methods is the inclusion of the decay channels (in the
case of the $\Delta$ the process $\Delta\to N \pi$).  This process is
not included in other methods, and yet is a vitally important and
physically based consideration.  However, because of the finite nature of
the lattice, decays are not always possible.  The finite periodic
volume of the lattice restricts the available momenta to discrete
values
\be
k_{\mu} = \frac{2\pi n_{\mu}}{aL_{\mu}}, 
\hspace{4mm} {\rm with} \hspace{2mm}
-\frac{L_{\mu}}{2} < n_{\mu} \leq \frac{L_{\mu}}{2}
\ee
where $L_{\mu}$ and $a$ are the lattice size and
spacing in the $\mu$ direction, respectively.

\begin{figure}[htb]
\resizebox*{\columnwidth}{!}{\rotatebox{90}{\includegraphics{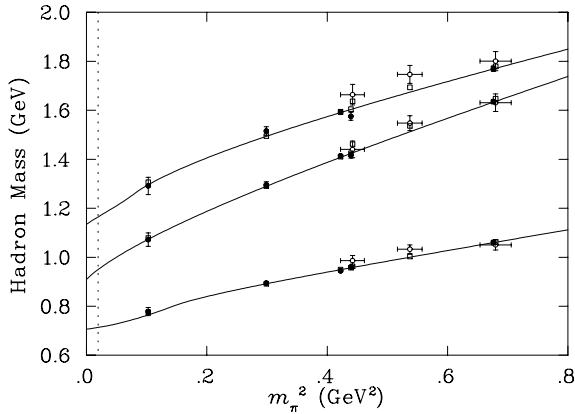}}}
\caption{Two-flavour, dynamical fermion lattice QCD data for the
\protect\( \Delta\protect \), \protect\( N\protect \) and vector meson
(\protect\( \rho\protect \)) mass
data from UKQCD \cite{Allton:1999gi} (open circles) and CP-PACS
\cite{Aoki:1999ff} (filled circles). The solid lines are the continuum
limit, infinite volume predictions of Eqs.~(\ref{eqn:Delta}),
(\ref{eqn:N}) and (\ref{eqn:rho}). The squares (barely discernable
from the data) are the predicted masses on a lattice of the same
dimensions as the data at that pion mass.\label{fig:Masses}}
\end{figure}

\begin{figure}[htb]
\resizebox*{\columnwidth}{!}{\rotatebox{90}{\includegraphics{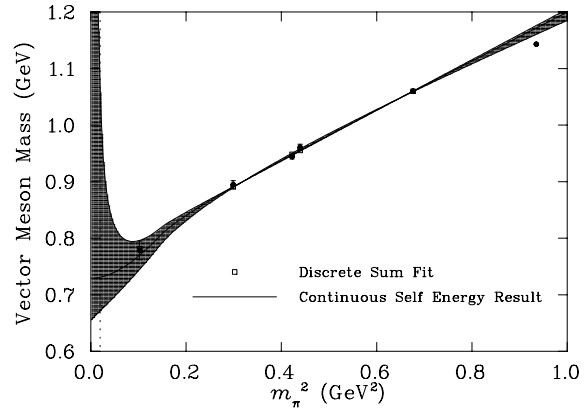}}}
\caption{Analysis of the lattice data for the vector meson (\protect\( \rho \protect \))
mass calculated by CP-PACS \cite{Aoki:1999ff} as a function of
\protect\( m_{\pi }^{2}\protect \). The shaded
area is bounded below by a 1\protect\( \sigma \protect \) error
bar. The upper bound is limited by a physical constraint discussed in
\cite{Leinweber:2001ac}.\label{fig:RhoErrors}} 
\end{figure}

Figure~\ref{fig:Masses} indicates the expected behaviour of the masses
of the $N$, $\Delta$ and $\rho$ using Eqs.~(\ref{eqn:N}),
(\ref{eqn:Delta}) and (\ref{eqn:rho}), with the physical masses being
$0.940$, $1.173$, $0.713$ GeV respectively.\footnote{The excellent agreement
with the experimental mass of the nucleon is coincidental.}  We also
present an error analysis of the fitting for the particular case of
the $\rho$ meson in Fig.~\ref{fig:RhoErrors}.  The shaded region is
bounded below by an increase of $1\sigma$ in the $\chi^{2}$ per degree
of freedom of the fit, and above by a physical constraint in our
approach.  It is clear that whilst the central value of the
extrapolation gives an acceptable value for the physical mass, the
uncertainties are large.  A Gedanken experiment performed in
\cite{Leinweber:2001ac} suggests that a ten-fold increase in the
number of configurations at the lowest pion mass data point
($\mpi^{2}\sim 0.1$ GeV$^{2}$) would reduce the uncertainty in the
extrapolated value to the 5\% level.

\section{OTHER QUANTITIES}

The advantage of calculating the mass of the hadrons in the manner
described above is that the form allows the direct extraction of other
properties of the hadron that depend upon the quark mass dependence of
the hadron mass.

\subsection{The Sigma Commutator}

The sigma commutator is a direct source of information about chiral
symmetry breaking within QCD\@ \cite{Sainio:2001bq}.  As such it is a
quantity of considerable importance to extract from lattice QCD
calculations.  The form of the commutator is
\bea
\sigma_{N} & = & \bar{m} \langle N | \bar{u}u + \bar{d}d | N \rangle
\label{eqn:SigmaComm:Matrix_Element} \\
           & = & \bar{m} \frac{\partial m_{N}}{\partial \bar{m}} \, ,
\label{eqn:SigmaComm:deriv}
\eea
where $\bar{m}$ is the average mass of the {\it up\/} and {\it down\/}
quarks.

$\sigma_{N}$ is not directly accessible via experiment, however world
data suggests a value of $45 \pm 8$ MeV \cite{Gasser:1991ce}.  Early
attempts at evaluating Eq.~(\ref{eqn:SigmaComm:deriv}) found results
in the range $15$ to $25$ MeV, and the attention soon changed to
evaluating the matrix element,
Eq.~(\ref{eqn:SigmaComm:Matrix_Element}), directly.  In quenched
calculations the results were in the 
$40$--$60$ MeV range, but a two flavour dynamical fermion calculation
by the SESAM collaboration \cite{Gusken:1999wy} found a value of $18
\pm 5$ MeV.  The difficulties associated with these approaches are
two-fold.  Firstly, the scale independent quantity of $\sigma_{N}$
must be constructed from the renormalisation depended quantities
$\bar{m}$ and $\langle N | \bar{u}u + \bar{d}d | N \rangle$.
Additionally there still is the need to extrapolate the quantities to
the physical pion mass.

Our recent work showed that provided the extrapolation method is under
control the evaluation of $\sigma_{N}$ at $\mpi = 140$ MeV, is a
straightforward calculation.  The important advantage of this approach
is that one need only work with renormalisation group invariant
quantities.

We discussed previously how a chirally motivated form, Eq.\
(\ref{eqn:cubic}), will not reproduce the lattice data if the coefficient
of the $\mpi^{3}$ term is the value required by $\chi$PT\@.  However we
also showed that allowing this coefficient to be a fit parameter results
in a value that is wrong by almost an order of magnitude.  This
becomes even more significant in the case of the sigma commutator. The
required derivative promotes this coefficient to greater
significance {\em and}\/ the sign of the terms acts to reduce the value of
$\sigma_{N}$.  However this is not an issue with the extrapolation
forms discussed above.  The sign and magnitude of the cubic term is
exactly that predicted by $\chi$PT, but the effects are countered by
higher order terms --- resulting in a prediction for the value of
$\sigma_{N}$ that included the correct chiral physics.  We find
\cite{Leinweber:2000sa} that the value of the sigma commutator is
approximately $45$ MeV.

\subsection{The $J$ Parameter}

\begin{figure}[htb]
\resizebox*{\columnwidth}{!}{\rotatebox{90}{\includegraphics{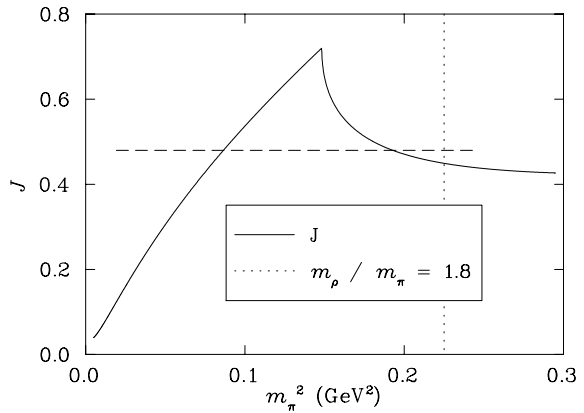}}}
\caption{The solid curve is a plot of the value of the \protect\( J\protect \)-parameter
as a function of \protect\( m_{\pi }^{2}\protect \) obtained from Eq.\
(\ref{eqn:J_param}) and the best fit to the lattice results.
The vertical dotted line shows the
point at which the \protect\( J\protect \)-parameter is evaluated
(\protect\( m_{\rho }/m_{\pi }=1.8\protect \)). The horizontal line
displays the experimental value (0.48) plotted between the physical
values of \protect\( m_{\pi }^{2}\protect \) and \protect\(
m_{K}^{2}\protect \).\label{fig:J_vs_mpi2}}
\end{figure}

This dimensionless parameter was proposed as a quantitative measure,
independent of the need for extrapolation --- an ideal lattice
observable \cite{Lacock:1995tq}.  It has the form
\bea
J & = & m_{\rho} \left. \frac{dm_{\rho}}{d\mpi^{2}}
        \right|_{m_{\rho}/\mpi = 1.8} \label{eqn:J_param}\\
  & \simeq & m_{K^{*}}\frac{m_{K^{*}} - m_{\rho}}{m_{K}^{2} -
             \mpi^{2}} \, ,
\eea
which, by substituting the experimental mass values, yields the value
\cite{Lacock:1995tq}
\bes
J = 0.48(2) \, ,
\ees
In Fig.~\ref{fig:J_vs_mpi2} we present the value of the $J$ parameter as
obtained from Eqs.~(\ref{eqn:rho}) and (\ref{eqn:J_param}).
The detailed slope of the curve is parameter dependent, however the
presence of the cusp is model independent.  The cusp is a result of
the two pion cut in the rho spectral function and has been ignored in
previous attempts at evaluating the $J$ parameter.  We find a value
for the $J$ parameter of $45(7)$ in good agreement with the
experimental value.  We note, however, that if the point of evaluation
corresponded to $\mpi^{2} \sim 0.15$ GeV$^{2}$ the $J$ parameter would
have been around $50$\% larger.

\subsection{Edinburgh Plot}

The baryon and meson masses on the lattice are all determined modulo the
lattice spacing -- a scale that must be determined from some piece of
data external to the lattice.
One method of removing this scale is by
plotting a ratio of masses --- the Edinburgh plot.
\begin{figure}[htb]
\resizebox*{\columnwidth}{!}{\rotatebox{90}{\includegraphics{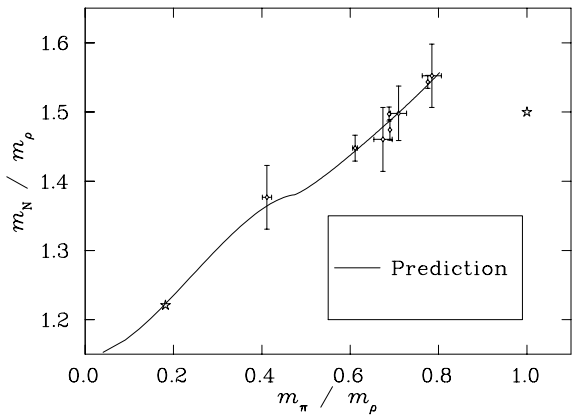}}}
\caption{Edinburgh plot for CP-PACS (filled symbols) and UKQCD (open
symbols) results. The stars represent the known limiting cases, at the
physical and heavy quark limits respectively. The solid line is the
infinite volume, continuum limit behaviour predicted by our functional
forms for the extrapolation of the \protect\( N\protect \) and
\protect\( \rho \protect \) masses.\label{fig:Edinburgh}}
\end{figure}
In Fig.~\ref{fig:Edinburgh} we present a prediction for the infinite
volume, continuum limit extrapolation of the lattice data previously
presented.  The two points known explicitly are indicated by open
stars on the plot.  The first known point is ratio of the physical
masses of the $\pi$, $\rho$ and $N$.  The second point is the heavy
quark limit, when the masses of the hadrons become proportional to the
constituent quarks.  The effect of the opening of the decay channel of
the rho is visible at $\mpi / m_{\rho} = 0.5$.  The effects
induced, and the expected behaviour on the finite sized lattice will
be presented in a future work \cite{Us:Ed}.

\section{SUMMARY}

The importance of including the correct chiral behaviour in
extrapolation methods is becoming more important as dynamical lattice
QCD results appear at lighter quark masses.  The successes of
the approach outlined above include not only predictions for the
physical masses of the hadrons investigated, but other quantities
successfully reproduced.  These other successes include the sigma
commutator and the $J$ parameter --- both of which have been a thorn
in the side of dynamical fermion calculations.  It is through the
inclusion of the dominant chiral physics, the recognition that decay
channels are important, and the understanding of some of the finite size
lattice artifacts that we have been able to successfully
extrapolate the Edinburgh plot to the known physical limit.

\section*{Acknowledgements}
This work was supported by the Australian Research Council and the
University of Adelaide.

\end{document}